# Relationship between EIT Post Eruption Arcades, Coronal Mass Ejections, Coronal Neutral Line and Magnetic Clouds

*Vasyl Yurchyshyn*
*Big Bear Solar Observatory,*
*40386 North Shore Lane, Big Bear City, CA 92314*

**Abstract.** There is observational evidence that the elongation of an Earth directed coronal mass ejection (CME) may indicate the orientation of the underlying erupting flux rope. In this study, we compare orientations of CMEs, MCs, eruption arcades and coronal neutral line (CNL). We report on good correlations between i) the directions of the axial field in EIT arcades and the elongations of halo CMEs and ii) the tilt of the CNL and MC axis orientations. We found that majority of the eruptions that had EIT arcades, CMEs and MCs similarly oriented also had the CNL co-aligned with them. To the contrary, those events that showed no agreement between orientations of the EIT arcades, CMEs and MCs, had their MCs aligned with the coronal neutral line. We speculate that the axis of the ejecta may be rotated in such a way that it is locally aligns itself with the heliospheric current sheet.

## 1. Introduction

Earth directed coronal mass ejections (CMEs), or halos, are known to be associated with various, often hazardous, disturbances in the near Earth environment. The key issue here is the presence in the ejecta of strong southward component of the magnetic field, which is known to be associated with geomagnetic storms [*Russell et al.*, 1974]. Better knowledge of CME's magnetic structure is, therefore, crucial for our advance in physical understanding and theoretical modeling of CME origin as well as space weather forecast.

CME are thought to represent a flux rope [*Moran and Davila*, 2004; *Krall and St. Cyr*, 2006; *Krall*, 2007] and are highly structured three-dimensional features [*Cremades and Bothmer*, 2004; *Krall and St. Cyr*, 2006]. The structured CMEs are though to be magnetically organized in the axial direction, which corresponds to the axis of a large scale twisted flux rope. A recent study [*Krall and St. Cyr*, 2006] showed that statistical parameters, measured for observed CMEs, such as eccentricity and the axial aspect ratio are in agreement with those obtained for a parameterized model flux rope.

White light coronagraphs, such as LASCO instrument on board SOHO observatory, show that halo CMEs have various sizes and shapes. Many of them can be enveloped by an ellipse and can be fitted with a cone model [*Zhao et al.*, 2002; *Xie et al.*, 2004; *Zhao*, 2004; *Michalek et al.*, 2005].

Erupting flux rope modeling [*Krall et al.*, 2006; *Yurchyshyn et al.*, 2006; *Yurchyshyn et al.*, 2007] showed that a model halo CMEs appear to be elongated in the direction of the underlying flux rope axial field. Therefore it is quite possible the ellipse shaped appearance of halo CMEs may be related to their magnetic structure.

Following this idea, we have measured the orientations of 25 ellipse shaped halos and the associated magnetic clouds (MCs) and reported that for about 64% of events, the difference between the orientations of halo elongations and MCs does not exceed ±45 deg [*Yurchyshyn et al.*, 2007, herefrom Paper I]. This finding was later confirmed by [*Zhao*, 2007].

This Letter is continuation of our research on the structure of erupting magnetic fields. Here we compare tilts of i) EUV post eruption arcades (PEA) usually associated with eruptions [*Tripathi et al.*, 2004], ii) halo CMEs, iii) heliospheric current sheet (HCS) near the eruption site and iv) axial field in MCs at 1AU, in order to further investigate CME evolution in the interplanetary space.

## 2. Method and Measurements

### 2.1 Orientations of halo CMEs

Our study is based on 25 events, selected from the Master Data Table compiled during a *Living With a Star* Coordinated Data Analysis Workshop and the list published in [*Qiu and Yurchyshyn*, 2005].

The CME orientation angle (tilt), $\alpha_{CME}$, was determined in Paper I by fitting an ellipse to an irregularly shaped "halo" around the C3 occulting disk (Figure 2 in Paper I) and measuring its tilt in the clockwise (CW) direction from the positive GSE *y*-axis to the ellipse semi-major axis. In the geocentric solar equatorial coordinate system (GSE), *y*-axis is in the ecliptic plane pointing towards dusk, *x*-axis is directed from the Earth towards the Sun and *z*-axis is pointed northward. Note, that we use the same coordinate system to determine all orientation angles discussed in this study. Because we measured 3-5 LASCO frames for each event, the final CME tilt, listed in Table I of Paper I, was calculated as the mean of all angles, determined from individual frames.

*2.2 Direction of the axial and azimuthal field in EUV post eruption arcades and CMEs*

In Paper I we did not take into account the possible direction of the axial field in a CME (i.e., underlying erupting flux rope), so that the CME orientations have 180 deg ambiguity in its direction. To remove the ambiguity we will use the magnetic field structure inferred from the associated PEA.

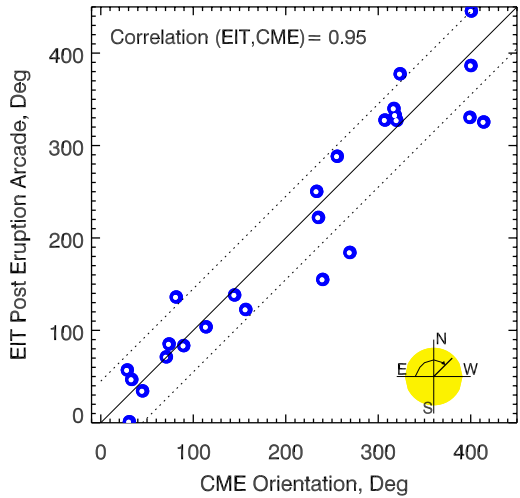

*Figure 1. Correlation between the directions of the axial field in CMEs and EIT post eruption arcades (PEA).*

The "standard" flare model [*Forbes*, 2000] predicts that eruption of a flux rope is accompanied by magnetic reconnection and formation of flare ribbons at the foot points of the PEA, which indicate the general orientation of the erupting flux rope. The CME ambiguity resolution is based on the assumptions that direction of the axial field and twist in a flux rope CME corresponds to those of the EIT/Hα flare arcade associated with the eruption. In general, the direction of the axial field and twist (helicity sign) in a PEA can be determined from solar data [see details in *Yurchyshyn et al.*, 2001]. We will require that the axial field of the PEA should make an acute angle with that of the CME, which can be achieved by adding 180 deg to the CME orientation angle.

In Figure 1 we show the correlation between the CME angles and PEA directions. For the overwhelming majority of events the difference between the angles is less that 45 deg. This indicates that, on average, the elongation of a halo CME is co-aligned with the EIT post eruption arcade and is additional evidence that the ellipse shape of a halo CME may indeed bear information on the geometry of the underlying flux rope. Because of this high correlation, for the comparisons with other parameters we will only use CME orientations and we will refer to them in the text below as PEA/CME orientations/angles.

*2.3 Orientation of the coronal neutral line as measured from coronal field maps*

The background in Figure 2 is the Wilcox Solar Observatory (WSO) coronal magnetic field map showing the polarity distribution (light and dark gray) during Carrington rotation (CR) 1968. Black solid line represents the major coronal neutral line (CNL). This map was calculated from a synoptic photospheric field map with a potential field model [*Hoeksema et al.*, 1983; *Hoeksema*, 1984]. The green diamond in Figure 2 indicates the location of the CME source region relative to the CNL, i.e., on the day when it crossed the central meridian. The averaged tilt of the neutral line near the eruption site was measured (in clockwise direction) as the tilt of a thick line segment centered on the point closest to the eruption site.

In order to compare directional angles of CMEs and MCs with the tilt of the CNL, which has 180 ambiguity in it, we needed to assign the direction to the CNL orientation. This was done the same way as we assigned the field direction in the CME: by requiring that the CNL directional angle makes an acute angle with the axial field of the corresponding MC. Our choice of MCs as a reference is justified as follows: i) both MCs and CNL are low-order, large-scale heliospheric structures, as opposed to the PEAs that represent high-order solar surface fields and ii) it is not necessary that each PEA is formed under the streamer belt, therefore, the PEA orientation may not always be related to the CNL.

*2.4 Magnetic cloud parameters*

For each event, the MC orientation angle was obtained by averaging the orientations produced with different MC fitting routines [*Yurchyshyn et al.*, 2007]. The orientation (clock) angle is the direction angle of the projected MC flux rope axis onto the GSE *yz* plane, measured in the clock-wise (CW) direction from the positive *y*-axis.

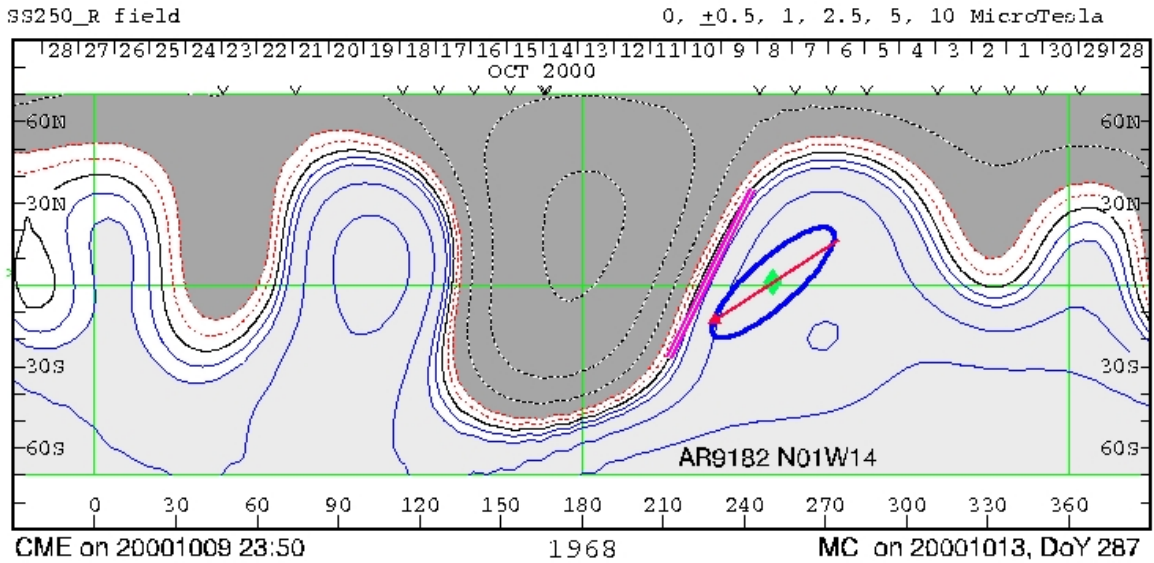

*Figure 2. Wilcox Solar Observatory coronal field map calculated for the height of 2.5 solar radii for CR 1968. Dashed red contours (dark grey) represent negative fields, blue contours (light grey) – positive fields. Thick solid line is the coronal neutral line (CNL). A red line segment shows the estimated orientation of the CNL near the site of eruption on October 13, 2000. The tilt of the CNL was measured in the CW direction from the east. The ellipse, centered at the location of the CME source (diamond) indicates the orientation of a halo CME.*

### 3. Results

In Figure 3 we present a correlation between the PEA/CME orientations and MC axes. The green symbols indicate those 15 (out of 25) events that showed a good correspondence between the PEA/CMEs and MCs directions (angle difference < 45 deg). The red symbols represent events where the angle difference exceeds 45 deg. Please, note that Figures 4 and 5 use the same color coding. Figure 4 plots PEA/CME directions versus the CNL tilt. Similarly to Figure 3, the green symbols in this plot are also mainly clustered around the bisector (solid line). All "red" events in Figure 4, except one, are located outside the ±45 deg range centered on the bisector. Figures 3 and 4 imply that those erupting flux ropes that where initially aligned with the CNL (streamer belt) at the early stage of eruption, appear to remain so when they reach the Earth, thus that their orientations match those of the MCs.

In Figure 5 we plot MC axis directions versus the tilt of the CNL near the eruption site. This graph displays a high correlation between the parameters with the data points tightly clustered around the bisector. The figure indicates that the MCs in our data set tend to be aligned along the neutral line. Considering that the CNL is the base of the heliospheric current sheet, it ultimately means that the MCs were embedded in the current sheet.

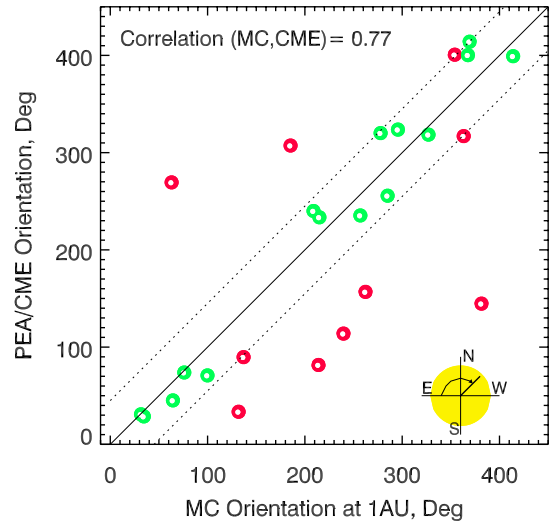

*Figure 3. PEA/CME orientation angles plotted vs the axial field directions in MCs. The green symbols indicate those events that showed a good correspondence (difference < 45 deg) between the two directional angles. Red symbols are the events that display no correspondence (difference > 45 deg).*

## 4. Conclusions and Discussion

First, we would like to briefly summarize our findings: i) there is a good correlation between the directions of the axial field in EIT post eruption arcades and the elongations of halo CMEs; ii) 80% of events in the data set display the difference between CNL and MC orientations to be less than 45 deg; iii) majority of the eruptions that had PEA/CMEs and MCs similarly oriented (i.e. "green" events) also had the magnetic neutral line co-aligned with them; iv) those PEA/CME-MC pairs that showed no agreement between the PEA/CME and MC orientations ("red" events), had their MCs aligned with the magnetic neutral line.

As we mentioned earlier, observations and theoretical works indicate that the coronal ejecta may evolve substantially as it expands out into heliosphere and interacts with heliospheric and solar wind magnetic fields. Shape, structure as well as magnetic field connectivity of the interplanetary ejecta may change due to interactions with ambient solar wind.

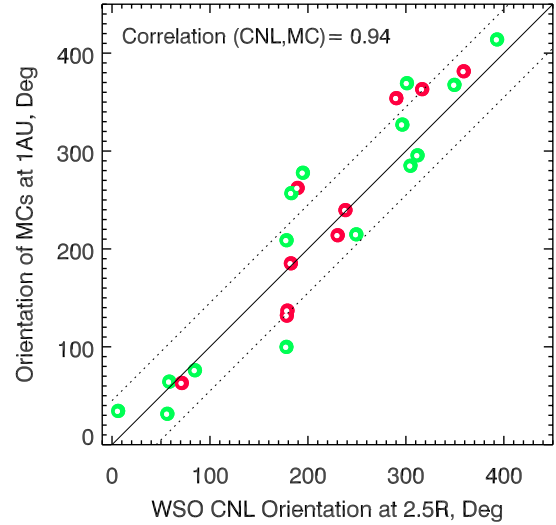

*Figure 5. Directions of MC axial fields plotted vs the orientation of the CNL. The green symbols indicate the same events as in Figure 3. Majority of green and red symbols in this plot are located at or near the bisector.*

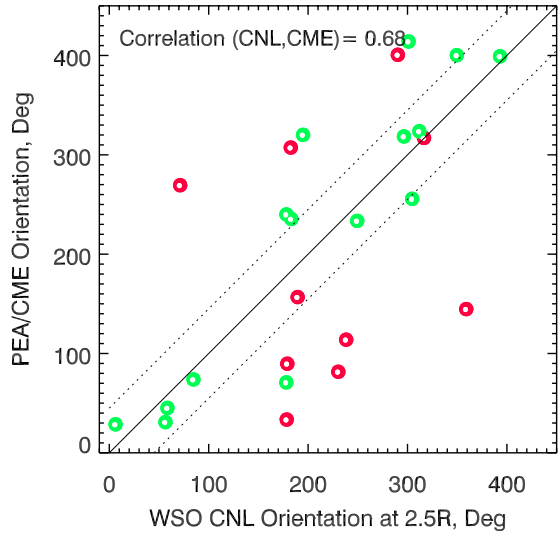

*Figure 4. PEA/CME orientation angles plotted vs the orientation of the CNL near the eruption site. The green/red symbols in this plot indicate the same events as in Figure 3. Note that majority of "green" symbols are also located at or near the bisector.*

The data presented here show that coronal mass ejections are significantly affected by the heliospheric magnetic fields: they have a tendency not only to be deflected toward the heliomagnetic equator and channeled into the HCS

[*Crooker et al.*, 1993; *Zhao and Hoeksema*, 1996; *Mulligan et al.*, 1998; *Kahler et al.*, 1999] but also the axis of the flux rope appears to be deformed in such a way that it locally aligns itself with the heliospheric current sheet. This conclusion, based on a detailed study of 25 events is in agreement with the earlier reports that i) MCs, oriented between ±30 deg, tend to be detected slightly more frequently [*Zhao and Hoeksema*, 1998] and ii) during solar minimum (maximum) dominate bipolar (unipolar) MCs [*Mulligan et al.*, 1998]. Note that bipolar (unipolar) MCs are nearly parallel (perpendicular) to the ecliptic.

To further check these inferences, we used coronal field models that show the shape and position of the neutral line at different heights above the solar surface. Those models were produced by the SAIC solar physics group [*Lionello et al.*, 2005]. The left panel in Figure 6 shows a coronal field map at 1.6 solar radii and the orientation of the halo CME (ellipse), while the right panel shows a coronal field map at 16.5 radii and the orientation of the corresponding MC (arrow). As it follows from the figure, the CME elongation initially matched the local tilt of the neutral line. However, further out from the Sun, the neutral line changes its orientation and, apparently, the interplanetary ejecta does the same, in such a way

5that the corresponding MC is well aligned with the neutral line. The present results seem to be consistent with the above indications that the heliospheric magnetic field may significantly influence coronal eruptions.

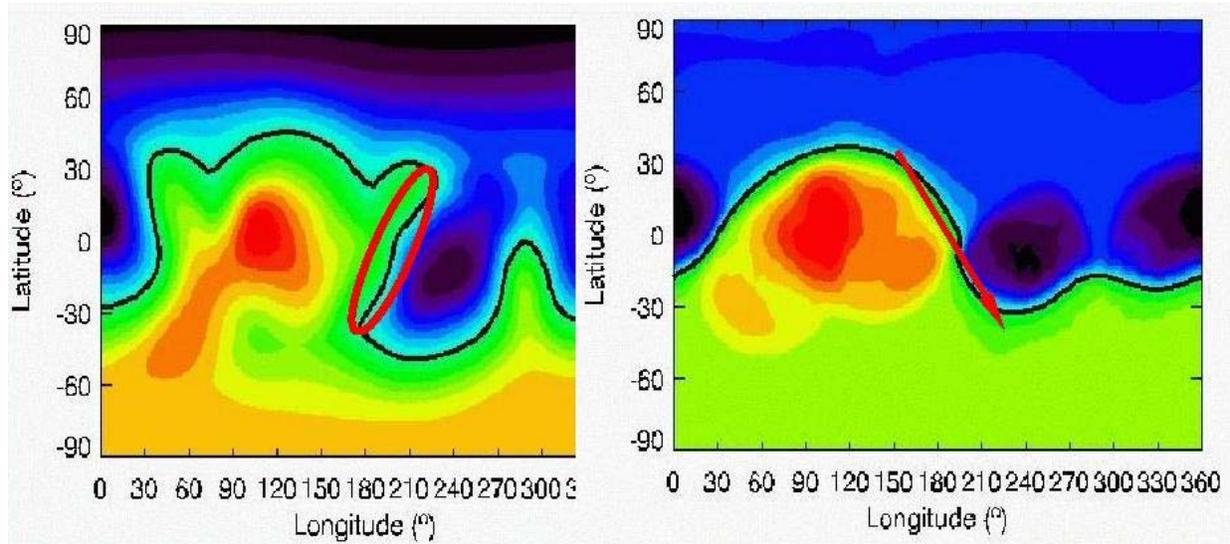

*Figure 6. Coronal field maps calculated at 1.6 (left) and 16.5 (right) solar radii for CR 2006 by SAIC solar physics group. The thick black contour is the coronal magnetic neutral line. The oval represents a halo CME on Aug 14, 2003, aligned along the CNL at 1.6 radii. The arrow in the right panel indicates the axis of the corresponding MC.*

**References**

Cremades, H., and V. Bothmer (2004), On the three-dimensional configuration of coronal mass ejections, *Astronomy and Astrophysics*, *422*, 307-322.

Crooker, N. U., et al. (1993), Multiple heliospheric current sheets and coronal streamer belt dynamics, *Journal of Geophysical Research*, *98*, 9371-9381.

Forbes, T. G. (2000), A review on the genesis of coronal mass ejections, *Journal of Geophysical Research*, *105*, 23153-23166.

Hoeksema, J. T., et al. (1983), The structure of the heliospheric current sheet - 1978-1982, *Journal of Geophysical Research*, *88*, 9910-9918.

Hoeksema, J. T. (1984), Structure and evolution of the large scale solar and heliospheric magnetic fields, 5 pp.

Kahler, S. W., et al. (1999), The polarities and locations of interplanetary coronal mass ejections in large interplanetary magnetic sectors, *Journal of Geophysical Research*, *104*, 9919-9924.

Krall, J., and O. C. St. Cyr (2006), Flux-Rope Coronal Mass Ejection Geometry and Its Relation to Observed Morphology, *Astrophysical Journal*, *652*, 1740-1746.

Krall, J., et al. (2006), Flux Rope Model of the 2003 October 28-30 Coronal Mass Ejection and Interplanetary Coronal Mass Ejection, *Astrophysical Journal*, *642*, 541-553.

Krall, J. (2007), Are All Coronal Mass Ejections Hollow Flux Ropes?, *Astrophysical Journal*, *657*, 559-566.

Lionello, R., et al. (2005), The Effects of Differential Rotation on the Magnetic Structure of the Solar Corona: Magnetohydrodynamic Simulations, *Astrophysical Journal*, *625*, 463-473.

Michalek, G., et al. (2005), Properties and geoeffectivenes of halo coronal mass ejections, *Space Weather*, *4*, S10003.

Moran, T. G., and J. M. Davila (2004), Three-Dimensional Polarimetric Imaging of Coronal Mass Ejections, *Science*, *305*, 66-71.

Mulligan, T., et al. (1998), Solar cycle evolution of the structure of magnetic clouds in the inner heliosphere, *Geophysical Research Letters*, *25*, 2959-2962.

Qiu, J., and V. B. Yurchyshyn (2005), Magnetic Reconnection Flux and Coronal Mass Ejection Velocity, *Astrophysical Journal*, *634*, L121-L124.

Russell, C. T., et al. (1974), On the cause of geomagnetic storms, *Journal of Geophysical Research*, *79*, 1105-1109.

Tripathi, D., et al. (2004), The basic characteristics of EUV post-eruptive arcades and their role as tracers of coronal mass ejection source regions, *Astronomy and Astrophysics*, *422*, 337-349.

Xie, H., et al. (2004), Cone model for halo CMEs: Application to space weather forecasting, *Journal of Geophysical Research (Space Physics)*, *109*, 03109.

Yurchyshyn, V., et al. (2006), The May 13, 2005 Eruption: Observations, Data Analysis and Interpretation, *Solar Physics*, *239*, 317-335.

Yurchyshyn, V. B., et al. (2001), Orientation of the Magnetic Fields in Interplanetary Flux Ropes and Solar Filaments, *Astrophysical Journal*, *563*, 381-388.